\documentclass[runningheads]{llncs}
\usepackage{graphicx}
\usepackage{soul}
\usepackage{wrapfig}
\usepackage{booktabs}
\usepackage{graphicx}
\usepackage[table,xcdraw]{xcolor}
\usepackage{xcolor}

\begin{document}
\title{Adversarial Text Purification: A Large Language Model Approach for Defense}
%
%
\author{Raha Moraffah\tocauthor{Raha~Moraffah} \and
Shubh Khandelwal \and
Amrita Bhattacharjee\tocauthor{Amrita~Bhattacharjee} \and
Huan Liu\tocauthor{Huan~Liu}}

\institute{Arizona State University, Tempe, AZ, USA  \\
\email{\{rmoraffa,skhand15,abhatt43,huanliu\}@asu.edu}}
\authorrunning{Moraffah et al.}
%
%
\maketitle              
\begin{abstract}

Adversarial purification is a defense mechanism for safeguarding classifiers against adversarial attacks without knowing the type of attacks or training of the classifier. These techniques characterize and eliminate adversarial perturbations from the attacked inputs, aiming to restore purified samples that retain similarity to the initially attacked ones and are correctly classified by the classifier. 
Due to the inherent challenges associated with characterizing noise perturbations for discrete inputs, adversarial text purification has been relatively unexplored.  In this paper, we investigate the effectiveness of adversarial purification methods in defending text classifiers. We propose a novel adversarial text purification that harnesses the generative capabilities of Large Language Models (LLMs) to purify adversarial text without the need to explicitly characterize the discrete noise perturbations. We utilize prompt engineering to exploit LLMs for recovering the purified samples for given adversarial examples such that they are semantically similar and correctly classified. Our proposed method demonstrates remarkable performance over various classifiers, improving their accuracy under the attack by over 65\% on average.

\end{abstract}
\section{Introduction}
%

Despite the tremendous success of text classification models~\cite{devlin2018bert}\cite{liu2019roberta}, studies have exposed their susceptibility to adversarial examples, i.e., carefully crafted sentences with human-unrecognizable changes to the inputs that are misclassified by the classifiers~\cite{jin2020bert}. The dependability and integrity of NLP applications are seriously threatened by the vulnerability of text classification models to these attacks.
Thus, developing stronger defenses against adversarial attacks is crucial in improving the classification model's robustness. 

Adversarial purification is a type of defense mechanism against adversarial attacks. It characterizes and removes the adversarial perturbations from the attacked inputs to generate purified samples that are similar to the attacked ones and are classified correctly by the classifier~\cite{nie2022diffusion}\cite{samangouei2018defense}\cite{yoon2021adversarial}\cite{shi2021online}. 
~These methods have demonstrated efficacy in the field of image classification without making assumptions on the form of an attack and a classification model, thus being able to defend pre-existing classifiers against unseen threats.  
The potential of adversarial purification, however, has not been explored for text classification, due to the challenges of characterizing the adversarial perturbations for discrete data. In particular, contrary to images, where perturbations can be generated based on continuous gradients, for text data, adversarial perturbations are generated by manipulating combinations of words in the input text~\cite{jin2020bert}. Therefore, identifying these perturbations is also a combinatorial problem. 

An ideal solution to adversarial purification for text is to generate the purified example without explicitly characterizing the noise perturbations. In an attempt to achieve this, Li et al.~\cite{li2022text} propose a greedy approach that randomly masks the adversarial examples and uses their reconstructed versions by the Masked Language Models (e.g., BERT~\cite{devlin2018bert}) as benign purified examples. However, due to its greedy nature, this defense can be ineffective for defending text classifiers.

The exponential growth of the sheer size of LLMs has expedited their generative applications in various fields~\cite{peng2023study}. To study the effectiveness of adversarial purification for texts, we investigate if LLMs can be exploited to directly generate the purified examples from their adversarial counterparts, eliminating the need for the characterization of adversarial perturbation. To this end, we utilize the generative power of instruction-based LLMs, particularly GPT-3.5, and design a prompt to exploit the contextual understanding and capacity of LLMs to recover purified samples. 

Compared to the greedy approach of selecting random combinations of tokens iteratively to remove adversarial perturbations, our proposed method exploits the comprehension and contextual understanding of LLMs to effectively reverse the adversarial perturbations, while utilizing their extensive generation power and capacity to produce cohesive, fluent texts.
Our method demonstrates the effective use of adversarial purification methods for text classification, improving the performance of the classifier under attack by over 65\%, and improving the performance of the existing text purification defense by over 25\% in most cases.  Our results open a new avenue for future research in textual adversarial defense based on purification. Our contributions are summarized as follows:
\begin{itemize}
    \item We study if it is possible to effectively implement the adversarial text purification defense for text.
    
    \item We are the first to utilize the contextual understanding and capacity of LLMs for effective text-based adversarial purification defense.

 \item We conduct extensive experiments on two state-of-the-art transformer-based text classifiers and demonstrate the effectiveness of our proposed adversarial purification method in defending the pre-trained classifiers against strong attacks without any knowledge of the attack.

\end{itemize}

\begin{wrapfigure}{R}{0.45\textwidth}
    \centering
    \includegraphics[width=0.4\textwidth]{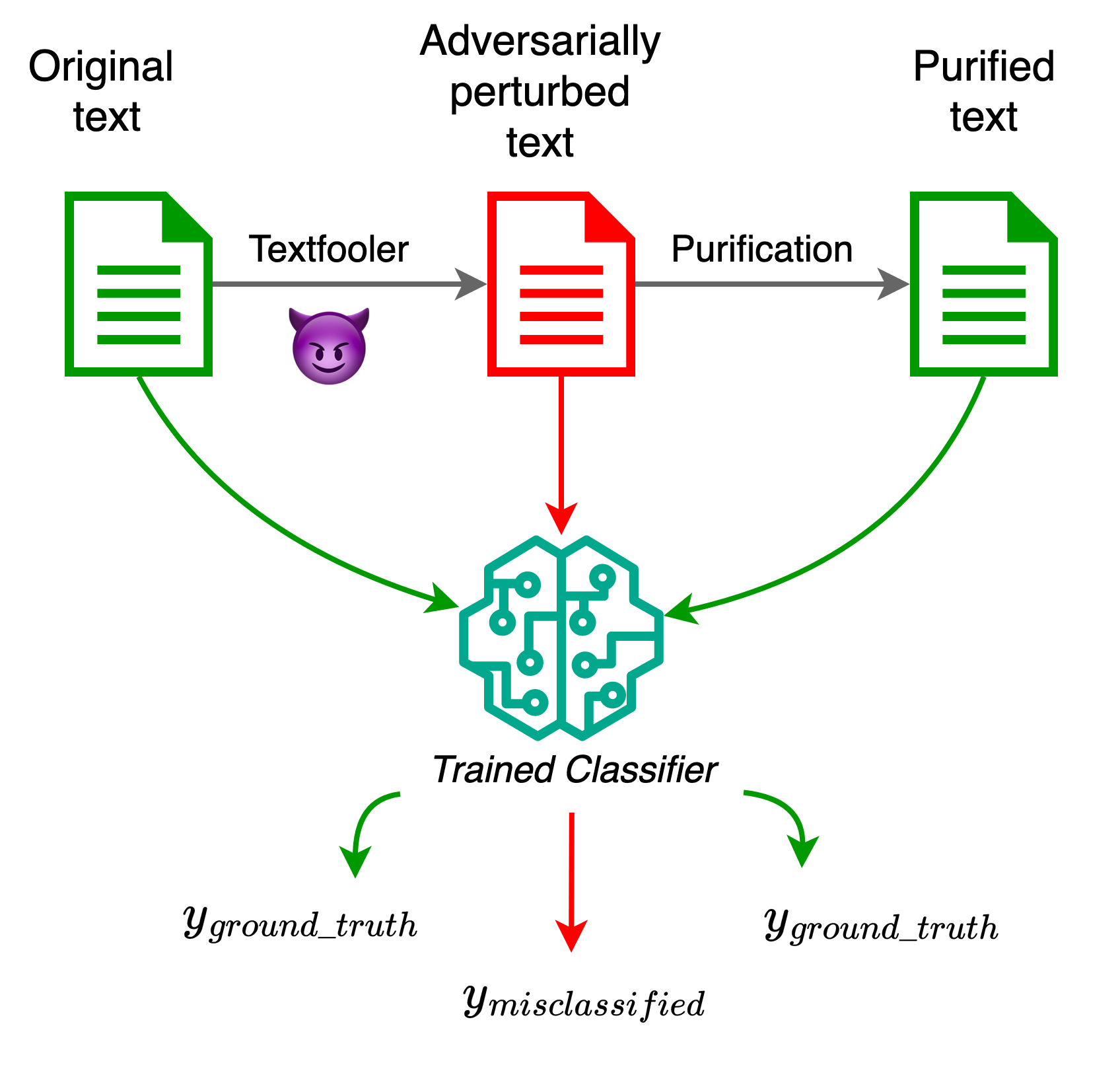}
    \caption{Our Proposed LLM-guided Adversarial Text Purification Framework.}
    \label{fig:framework}
\end{wrapfigure}
\section{Related Work} 

\textbf{Adversarial attacks on text classifiers: } Over the years there have been various types of adversarial attacks for text, with varying degrees of success on different types of model architectures. Adversarial attacks, broadly categorized into black box and white box ~\cite{shreya2022survey}, manipulate textual data through insertion, deletion, or swapping of characters and words. The substitution-based strategies to craft adversarial examples employ techniques like genetic algorithms, greedy-search, or gradient-based methods for word replacement~\cite{alzantot2018generating,jin2020bert,ren2019generating}. 
Recent works involving word-level perturbations include TextFooler ~\cite{jin2020bert}, BERT-Attack ~\cite{li2020bert}, TextHoaxer~\cite{ye2022texthoaxer}. Alongside the vast bidy of work on word-level attacks, there is also significant amount of works in character-level and sentence-level attacks~\cite{shreya2022survey}.

\textbf{Adversarial purification \& other defenses: } Influenced by the rapid development of various adversarial attacks in text, there has also been an increasing number of defense mechanisms to ensure robustness of models against different types of attacks. Some of these defense methods introduce certified robust models to create a defensive range within which substitutions cannot perturb the model ~\cite{jia2019certified}. Gradient-based adversarial training strategies have shown effectiveness in defending attacks with no prior knowledge and improving defense ~\cite{miyato2016adversarial,madry2017towards,ebrahimi2017hotflip,cheng2019robust,zhu2019freelb,li2021token}. Adversarial purification is a particularly desirable type of defense since it does not require prior knowledge of the type of attack. Prior work in adversarial purification has traditionally focused on continuous inputs ~\cite{li2022text} such as images, exploring generative models such as GANs~\cite{samangouei2018defense}, EBMs ~\cite{lecun2006tutorial}, and diffusion models ~\cite{song2020score,nie2022diffusion}. However, the field of creating better adversarial defenses and improving robustness in NLP has experienced considerable interest in recent years. Adversarial purification has been explored, however, it is comparably uncommon in NLP. ~\cite{li2022text} aims to utilize the contextual and masking capabilities of pre-trained masked language models (such as BERT ~\cite{devlin2018bert}) in order to create a defense against adversarial attacks. However, in this work, we aim to use the power of generative AI, in particular, recent state-of-the-art Large Language Models (LLMs) to perform adversarial purification in the context of capabilities to explore the possibility of improving the robustness of the models. 

\textbf{LLMs as pseudo-oracles:} Alongside the impressive performance of LLMs on a variety of natural language tasks~\cite{chang2023survey}, LLMs are also being increasingly used as pseudo-oracles, such as in data annotation ~\cite{flamholz2023large,alizadeh2023open,latif2023can}, as detectors~\cite{bhattacharjee2023fighting}, for model explainability~\cite{bhattacharjee2023llms} and as experts in general~\cite{xu2023expertprompting}. Inspired by such works, in this work, we propose to use LLMs to perform adversarial purification in the challenging text domain.

\section{Background}


\subsection{Large Language Models} 
Large language models (LLMs) are essentially deep networks that are based on transformer networks. Transformer-based LLMs are highly effective models that are capable of learning and generating natural language. Broadly there are two categories of language models: (i) Autoregressive language models and (ii) Masked language models. Autoregressive language models are simply trained to predict the next token in a sentence, thereby learning how to generate fluent text when pre-trained on a large corpora of data. Such models include GPT-2~\cite{radford2019language}, GPT-3~\cite{brown2020language}, etc. Masked language models (MLMs) are bi-directional models that learn by first masking some fraction of tokens in the sentence and then predicting appropriate tokens to fill the masked slots. Examples of such models include BERT~\cite{devlin2018bert}, RoBERTa~\cite{liu2019roberta}, etc. The bidirectional nature of MLMs help the models to have higher language understanding capabilities, and thereby better performance on NLU tasks. More recently, autoregressive models such as the GPT-3 family of models are also being further trained via instruction-tuning~\cite{ouyang2022training} with (instruction, response) text pairs, whereby the model learns to generate text to follow user-specified instructions and perform tasks. Some of these instruction-tuned models undergo further training steps (e.g., via RLHF~\cite{bai2022training}) to align their responses with human preferences. State-of-the-art LLMs such as GPT-3.5 and GPT-4 from OpenAI demonstrate impressive performance when it comes to understanding long and complex human-written instructions in the prompts, as well as editing and generating text. Therefore, we use one of these models in an off-the-shelf manner for our framework.

\subsection{Adversarial Text Purification}

Adversarial purification is an adversarial defense mechanism that is relatively newer in the natural language domain. As we elaborated in the previous section, this method has been well explored in the domain of computer vision, whereby generative models are used to perform the purification. In the image domain, the standard method is to inject random noise into a perturbed input image, and then use a generative model i.e., the purification algorithm to reconstruct the original \textit{clean} image from the noisy image over multiple rounds. The generated image would now be free of the adversarial perturbations.  However, in the domain of text, the discrete nature of the input makes it infeasible to apply the standard computer vision methods directly. 
One recent attempt at adversarial text purification ~\cite{li2022text} uses masked language models to randomly mask multiple copies of the perturbed text, and then recovering the text by filling in the mask using the masked language model. This method essentially is somewhat similar to the standard process of injecting noise and iteratively reconstructing the input, as followed in the image domain. However, there is no other method for performing adversarial text purification. To fill this gap, we propose to directly leverage the instruction understanding and text generation capabilities of recent state-of-the-art LLMs and use these LLMs to perform the text purification.


\section{LLM-guided Adversarial Text Purification}
\label{sec:framework}

In this section, we describe our purification framework and explain the necessary design choices. 

We show our overall framework in Figure \ref{fig:framework}. As mentioned previously, in this work we focus on the task of text classification and we use fine-tuned pre-trained language models (such as BERT~\cite{devlin2018bert}), denoted by $f(\cdot)$ as the classifier. During inference, we evaluate such a classifier on the test set of our task dataset $(X_{test}, Y_{test})$ where $X_{test}$ and $Y_{test}$ are the sequence of input texts and associated ground truth labels respectively. For an input text $x_i \in X_{test}$, say the classifier correctly predicted $f(x_i) = y_{i}$, or $y_{ground\_truth}$ for ease of reference. Now, say this text is perturbed by an adversarial attack method such that the perturbed text $x_i'$ now gets misclassified to a different label, say $y_{misclassified}$. While many defense mechanisms train the model i.e., the classifier to be adversarially robust to some specific categories of perturbations, purification methods enable simply editing the text, ideally removing the adversarial perturbation from the text and thereby enabling the model to correctly classify the text. Following this, we collect this set of adversarially perturbed input texts $X_{test}'$ and attempt to purify them by using off-the-shelf large language models. In order to do this, we carefully design prompts, as elaborated in the following paragraph. After the purification step, we obtain $\tilde{X}_{test}$ which then is correctly classified by the classifier in majority of the cases.

We use an instruction-tuned LLM which is capable of following human-written instructions in the prompt, in order to generate the purified samples. To enable this, we carefully design the following prompt:

\vspace{1.5mm}
\noindent\fbox{%
    \parbox{0.95\columnwidth}{%
`Human: You are a teacher tasked with grading a quiz.
The quiz consists of a sentence (the question) and a classification label (the
student's answer).\\
Unfortunately, the sentence has been manipulated by an adversarial attack,
leading to a misclassification.\\
\colorbox{pink}{Given the altered sentence and its incorrect label, your job is to generate} \colorbox{pink}{a new
sentence that is semantically similar to the altered one but will} \colorbox{pink}{be classified
correctly according to the correct label.}\\
The categories for classification are: \textcolor{blue}{[list of classification categories]}\\
\colorbox{pink}{ALTERED SENTENCE (QUESTION): \textcolor{blue}{[altered sentence]}}\\
\colorbox{pink}{MISCLASSIFIED LABEL (STUDENT ANSWER): \textcolor{blue}{[misclassified label]}}\\
\colorbox{pink}{CORRECT LABEL (TRUE ANSWER): \textcolor{blue}{[correct label]}}\\
Please create a new sentence that conveys the same meaning as the altered sentence \colorbox{pink}{but will be classified under the CORRECT LABEL when} \colorbox{pink}{graded}.\\
\colorbox{pink}{Even if there is not a misclassification, provide/construct the sentence to} \colorbox{pink}{the
best of your capability.}
The output format must be json:\\
{``Original Sentence": ``[New sentence here]"}
Begin!'

    }%
}
\vspace{1.5mm}

In the prompt above, \textcolor{blue}{[altered sentence]} refers to the adversarially perturbed input text $x_i'$, \textcolor{blue}{[misclassified label]} refers to $y_{misclassified}$, \textcolor{blue}{[correct label]} refers to $y_{ground_truth}$ and \textcolor{blue}{[list of classification categories]} refer to the list of possible labels for the particular classification task. As evident in the prompt, we `prime' the LLM to enable it to act like a knowledgable teacher, thereby guiding the editing process. This is the prompt we use for eliciting the purified version of the text from the LLM, and we denote this prompt as \texttt{P0}.

To investigate the efficacy of this carefully designed prompt, we further design and test out two variants of this prompt: \texttt{P1}: which removes the instruction regarding generating text that would correct the misclassified label, and \texttt{P2}: which essentially prompts the LLM to generate a paraphrased version of the input text. The prompt \texttt{P1} is created by simply removing the text highlighted in \colorbox{pink}{pink} from \texttt{P0}. Finally, the prompt \texttt{P2} is:



\vspace{1.5mm}
\noindent\fbox{%
    \parbox{0.95\columnwidth}{%
`Human: Please generate a paraphrased sentence version of the following sentence.\\
SENTENCE: \textcolor{blue}{[altered sentence]}\\
The output format must be json:\\
{``Original Sentence": ``[Paraphrased sentence here]"}
Begin!'

    }%
}
\vspace{1.5mm}

\section{Experiments}
We conduct comprehensive experiments to evaluate the effectiveness of our proposed LLM-guided adversarial purification method. Our experiments are designed to examine the three main aspects of our method: \textbf{(i)} Effectiveness of the proposed method;\textbf{(ii)} Ablation study of the components of the designed prompt; and, \textbf{(iii)} case study of the purified examples. In the following, we first explain our experimental setting and then discuss our experimental results.
\subsection{Experimental Setting}
\label{methods}

In this section, we describe the datasets, the adversarial attack and the LLM we used in our experiments. We also describe the relevant defense baselines we compare our method to and provide information on our experimental setup to ensure reproducibility. Note that our experimental settings closely follow the ones in the state-of-the-art methods~\cite{li2022text}.

\textbf{Datasets.} We conduct experiments on two commonly-used benchmark NLP datasets: (1) \textbf{IMDb}~\cite{maas2011learning}: for sentiment classification of movie reviews where each review is labeled with a \textit{positive} or \textit{negative} label, and (2) \textbf{AG News}~\cite{zeng2023certified}: news topic classification where each article is labeled with one of the four categories of \{\textit{science, business, world, sports}\}.\\
\begin{table}[ht]
\centering
\begin{tabular}{@{}cccc@{}}
\toprule
\multicolumn{1}{l}{\textbf{Defense} $\downarrow$} &
  \cellcolor[HTML]{D9EAD3}\begin{tabular}[c]{@{}c@{}}Original \\ Accuracy\end{tabular} &
  \cellcolor[HTML]{F4CCCC}\begin{tabular}[c]{@{}c@{}}TextFooler\\  ($K=12$)\end{tabular} &
  \cellcolor[HTML]{F4CCCC}\begin{tabular}[c]{@{}c@{}}TextFooler\\  ($K=50$)\end{tabular} \\ \midrule \midrule
\multicolumn{1}{l}{\textbf{IMDb} $\downarrow$}    & \multicolumn{1}{l}{} & \multicolumn{1}{l}{} & \multicolumn{1}{l}{} \\ \midrule 
\rowcolor[HTML]{CFE2F3} 
fine-tuned BERT                    & 94.1                 & 20.4                 & 2.8                  \\ \midrule
Adv-HotFlip (BERT)                 & 95.1                 & 36.1                 & 8.0                    \\
FreeLB (BERT)                      & 96.0                   & 30.2                 & 7.3                  \\
FreeLB++ (BERT)                    & 93.2                 & -                    & 45.3                 \\
Text purification (BERT) ~\cite{li2022text}    & 93.0                   & \underline{81.5}                    & 51.0                 \\ 
Text purification (RoBERTa) ~\cite{li2022text}    & 96.1                   & \textbf{84.2}                    & 54.3                 \\ \midrule
(\textit{Ours}) LLM-guided purification (BERT)     & 94.54 & 79.34 & \underline{73.52} \\ 
(\textit{Ours}) LLM-guided purification (RoBERTa)     & 95.06 & 78.9 & \textbf{76.16} \\ \midrule
\multicolumn{1}{l}{\textbf{AG News} $\downarrow$} & \multicolumn{1}{l}{} & \multicolumn{1}{l}{} & \multicolumn{1}{l}{} \\ \midrule
\rowcolor[HTML]{CFE2F3} 
fine-tuned BERT                    & 92.0                   & 32.8                 & 19.4                 \\ \midrule
Adv-HotFlip (BERT)                 & 91.2                 & 35.3                 & 18.2                 \\
FreeLB (BERT)                      & 90.5                 & 40.1                 & 20.1                 \\
Text purification (BERT) ~\cite{li2022text}    & 90.6                 & 61.5                 & 34.9                 \\ 
Text purification (RoBERTa) ~\cite{li2022text}    & 90.8                 & 59.1                 & 34.2                 \\ \midrule
(\textit{Ours}) LLM-guided purification (BERT)    & 95.12 & \textbf{83.58} & \underline{81.3} \\ 
(\textit{Ours}) LLM-guided purification (RoBERTa)     & 94.76 & \underline{82.84} & \textbf{81.4} \\ \bottomrule
\end{tabular}
\caption{Comparison of our LLM-guided purification methods with baselines as described in Section \ref{methods}. Post-attack accuracy numbers are as reported in ~\cite{li2022text}. \textbf{Bold} denotes the best performance in terms of recovered accuracy, and \underline{underline} implies the second-best performance.}
\label{tab:main-result}
\end{table}
~~\textbf{Adversarial Attack and Defense Baselines.} For all our experiments we use the one of the strongest textual attacks named TextFooler~\cite{jin2020bert}. Similar to our baselines, we use the open-source implementation of TextAttack library~\cite{morris2020textattack}. The TextFooler attack is selected due to its efficient generation of strong and highly successful adversarial examples, making it an ideal attack to assess the effectiveness of the defense mechanisms. Following previous work, for the size of candidate list we choose $K=\{12,50\}$ in our experiments.

Following the previous work on adversarial purification~\cite{li2022text}, we compare the performance of our method with two types of adversarial defense, namely (1) \textit{Textual adversarial training methods:} these methods are based on adversarial training of the classifiers using the adversarial examples generated based on the gradients of the latent space. We use \textbf{Adv-HotFlip~\cite{ebrahimi2017hotflip}} and \textbf{FreeLB~\cite{li2021token}}, two state-of-the-arts in this category that do not require   
For the choice of baseline defenses, as well as the \textbf{FreeLB++~\cite{li2021searching}}, which requires the candidate list; and (2) \textit{Textual adversarial purification methods:} methods based on purifying the adversarial examples to generate correctly-classified benign examples. To the best of our knowledge, only one text adversarial purification method exists as in~\cite{li2022text}. We include this method as our baseline. 




\textbf{Classifier.} Following work in ~\cite{li2022text} we use classifiers based on two pre-trained masked language models: BERT~\cite{devlin2018bert} and RoBERTa ~\cite{liu2019roberta}. For each dataset, we use BERT and RoBERTa models from Huggingface Transformers (bert-base-uncased\footnote{https://huggingface.co/bert-base-uncased} and roberta-base\footnote{https://huggingface.co/roberta-base}), fine-tuned on that specific dataset. Note that our proposed method does not require any further fine-tuning or adversarial training of the model and we can simple query the fine-tuned BERT and RoBERTa models in an off-the-shelf manner. For evaluating our framework, we report the post-attack accuracy, with and without the purification method, along with the original classifier accuracy without any attack.

\textbf{Implementation details}. We use OpenAI's GPT-3.5 (version as of November 2023) and use our carefully designed prompts to obtain purified versions of adversarially altered texts. The process involved crafting prompts that guide the model to generate semantically similar but unperturbed versions of the input texts. We chose GPT-3.5 for its advanced contextual understanding and generative capabilities as indicated in ~\cite{brown2020language}. We automated this process using the OpenAI API\footnote{https://platform.openai.com/docs/api-reference} and LangChain\footnote{https://www.langchain.com/}.
Our experiments were implemented in Pytorch and were run on two systems: (1) Linux system with one A30 and (ii) Linux system with four A100s. All code and links to data will be made available.

\subsubsection{Effectiveness of Proposed Purification Method}
\begin{wraptable}{R}{0.39\textwidth}
\centering
\begin{tabular}{@{}cc@{}}
\toprule
\begin{tabular}[c]{@{}c@{}}\textbf{Prompt} \\ \textbf{Type}\end{tabular} & \textbf{AG News} \\ \midrule \midrule
Original (BERT)                                        & 95.12   \\ \midrule
Full prompt \texttt{P0}                                         & \textbf{81.3}    \\
\texttt{P1}                                                     & 78      \\
\texttt{P2}                                                     & 52.7    \\ \bottomrule
\end{tabular}
\caption{Effectiveness of our full prompt as described in Section \ref{sec:framework} (denoted by \texttt{P0}).}
\label{tab:ablation}
\end{wraptable}




\subsection{Results \& Discussion}
In this section, we aim to answer if our proposed LLM-based adversarial text purification method is able to effectively purify the adversarial examples. For the sake of comparison, we also report the accuracy under attack for vanilla fine-tuned classifiers. We apply our defense and the state-of-the-art adversarial defenses on the IMDB and AG News datasets and report the results in Table~\ref{tab:main-result}. Our results demonstrate that our proposed method effectively defends the state-of-the-art transform-based text classifiers, improving their accuracy under attack by more than 60\% in most cases. We elaborate on our observations in the following: (1) The adversarial training-based defenses, i.e., Adv-HotFlip, FreeLB, and FreeLB++, are constantly outperformed by our method based on purification by a large margin (more than 30\%). This is because these models are robustified against continuous gradient-based adversarial perturbations and not the discrete word-level perturbations used by text adversarial attacks; (2) the state-of-the-art purification-based defense, namely Text purification, has remarkably lower performance compared to our method. This is because the Text purification method is based on a greedy approach and iteratively selects and perturbs random words. Our method, on the other hand, utilizes the power of LLMs to directly generate purified examples; and (3) finally, our proposed method (LLM-guided purification) achieves the highest after attack accuracy, which is comparable to the accuracy of the model before the attack. For instance, for the BERT trained on the AG News dataset, the original accuracy before the attack is 95.06\%, whereas the accuracy after the attack is 83.58\%, which is more that 20\% better than the accuracy under attack for the second best-performing defense (Text purification (BERT)).   

\begin{table}[ht]
\centering
\resizebox{\textwidth}{!}{%
\begin{tabular}{@{}ccc@{}}
\toprule
\multicolumn{1}{l}{} &
  \textbf{Texts} &
  \textbf{Label} \\ \midrule
\textit{Original} &
  \begin{tabular}[c]{@{}c@{}}E-mail scam targets police chief Wiltshire Police warns \\ about ``phishing" after its fraud squad chief was targeted.\end{tabular} &
  \textbf{\textcolor{blue}{\texttt{science}}} \\ \cmidrule(lr){2-2}
\begin{tabular}[c]{@{}c@{}}\textit{Adv.} \\ \textit{Perturbed} \\ ($K=12$)\end{tabular} &
  \begin{tabular}[c]{@{}c@{}}E-mail scam targets \textcolor{teal}{gendarmerie} chief Wiltshire \\ Police warns about ``phishing" after \\ its \textcolor{teal}{deception battalion massa} was targeted.\end{tabular} &
  \textbf{\textcolor{red}{\texttt{the world}}} \\ \cmidrule(lr){2-2}
\textit{LLM-purified} &
  \begin{tabular}[c]{@{}c@{}}Wiltshire Police issues warning about phishing \\ email scam targeting their deception battalion massa.\end{tabular} &
  \begin{tabular}[c]{@{}c@{}}\textbf{\textcolor{blue}{\texttt{science}}} \\ (conf.: 0.994)\end{tabular} \\ \cmidrule(lr){2-2}
\begin{tabular}[c]{@{}c@{}}\textit{Adv.} \\ \textit{Perturbed} \\ ($K=50$)\end{tabular} &
  \begin{tabular}[c]{@{}c@{}}E-mail scam targets police chief Wiltshire Police warns about ``phishing" \\ after its \textcolor{teal}{hoax battalion leiter} was targeted.\end{tabular} &
  \textbf{\textcolor{red}{\texttt{the world}}} \\ \cmidrule(lr){2-2}
\textit{LLM-purified} &
  \begin{tabular}[c]{@{}c@{}}Wiltshire Police alerts about a scam email targeting their police chief, \\ warning about phishing after their hoax battalion leiter was targeted.\end{tabular} &
  \begin{tabular}[c]{@{}c@{}}\textbf{\textcolor{blue}{\texttt{science}}} \\ (conf.: 0.984)\end{tabular} \\ \midrule
\textit{Original} &
  \begin{tabular}[c]{@{}c@{}}Consumer Prices Down, Industry Output Up  \\ WASHINGTON (Reuters) - U.S. consumer prices dropped in July  for the first time \\ in eight months as a sharp run up in energy  costs reversed, the government \\ said in a report that suggested  a slow rate of interest rate hikes is likely.\end{tabular} &
  \textbf{\textcolor{blue}{\texttt{business}}} \\ \cmidrule(lr){2-2}
\begin{tabular}[c]{@{}c@{}}\textit{Adv.} \\ \textit{Perturbed} \\ ($K=12$)\end{tabular} &
  \begin{tabular}[c]{@{}c@{}}\textcolor{teal}{Eaters Pricing} Down, \textcolor{teal}{Departments Product Arriba}  \\ WASHINGTON (Reuters) - U.S. \textcolor{teal}{consuming} prices \textcolor{teal}{declined} in July  \\ for the first time in eight months as a \textcolor{teal}{ferocious manage} up in energy \\  costs \textcolor{teal}{quashed}, the government tell in a \textcolor{teal}{notification that recommendations a} \\ \textcolor{teal}{sluggish cadence of relevance pace hiking is possible.}\end{tabular} &
  \textbf{\textcolor{red}{\texttt{science}}} \\ \cmidrule(lr){2-2}
\textit{LLM-purified} &
  \begin{tabular}[c]{@{}c@{}}Consumers face lower prices as government report suggests \\ slower pace of interest rate hikes due to decrease in energy costs.\end{tabular} &
  \begin{tabular}[c]{@{}c@{}}\textbf{\textcolor{blue}{\texttt{business}}} \\ (conf.: 0.954)\end{tabular} \\ \cmidrule(lr){2-2}
\begin{tabular}[c]{@{}c@{}}\textit{Adv.} \\ \textit{Perturbed} \\ ($K=50$)\end{tabular} &
  \begin{tabular}[c]{@{}c@{}}\textcolor{teal}{User Charging} Down, Industry \textcolor{teal}{Product} Up  \\ WASHINGTON (Reuters) - U.S. \textcolor{teal}{clients} prices dwindled  in July  for the first time \\ in eight months as a sharp run up in energy  costs \textcolor{teal}{quashed}, the government tell \\ in a report that \textcolor{teal}{recommendation}  a slow rate of interest rate hikes is likely.\end{tabular} &
  \textbf{\textcolor{red}{\texttt{science}}} \\ \cmidrule(lr){2-2}
\textit{LLM-purified} &
  \begin{tabular}[c]{@{}c@{}}U.S. consumer prices fell in July for the first time in eight months due to a \\ significant increase in energy costs, as reported by the government. \\ This suggests that the pace of interest rate hikes is likely to slow down.\end{tabular} &
  \begin{tabular}[c]{@{}c@{}}\textbf{\textcolor{blue}{\texttt{business}}} \\ (conf.: 0.999)\end{tabular} \\ \bottomrule
\end{tabular}%
}
\caption{Examples from the AG News dataset with TextFooler perturbations (with both $K=12$ and $K=50$) along with LLM-purified versions of the perturbed input. Portions of the input text altered by the TextFooler method are shown in \textcolor{teal}{teal}. Labels in \textcolor{blue}{blue} are correctly classified, labels in \textcolor{red}{red} are misclassified. We see that our methods successfully retains the original label after attack, while maintaining semantics of the original input.}
\label{tab:egs}
\end{table}

\subsubsection{Ablation: Effectiveness of Prompt Components}

In this section, we conduct experiments with two additional prompts namely \texttt{P1} and \texttt{P2} as explained in Section~\ref{sec:framework}, and compare their results with the results obtained using the main prompt (\texttt{P0}). Specifically, \texttt{P1} is designed to understand the effect of the explicit instruction to ensure the purified text is classified as the correct label. The goal of designing \texttt{P2} is to assess the effectiveness of our proposed prompt to ensure the purified samples retain semantic similarity to the adversarial counterparts. To this end \texttt{P2} simply asks the LLM to paraphrase the adversarial example. Our results reported in Table~\ref{tab:ablation} indicate the effectiveness of our main prompt. The accuracy under attack for purification based on \texttt{P1} is about 4\% less than the full prompt \texttt{P0}. This indicates that even though the full prompt is useful to achieve higher performance, our proposed methodology can obtain similar performance, even when the original correct label of the sample is unknown.  However, the performance achieved with \texttt{P2}is remarkably lower compared to the main prompt, indicating that our proposed prompt is indeed necessary for a successful adversarial purification.



\subsubsection{Case study}
We showcase some examples from the AG News dataset in Table \ref{tab:egs}. We can observe that our purified examples are semantically similar to the adversarial examples while being classified to the original  correct class before the attack. This shows that our method can successfully remove the adversarial perturbation and does not change the original benign content of the example. It is important to note that our method can effectively remove adversarial perturbations of any length with only one prompt. Additionally, our generated examples are fluent and grammatically correct, due to the generative power of the LLMs.










\section{Conclusion}
In this paper, we propose a novel text adversarial purification method, that can effectively remove the adversarial perturbations of any lengths from the adversarial examples and generate purified examples that are semantically similar but are classified to the original correct class. Overcoming the challenges of characterizing adversarial perturbations for discrete inputs (i.e., text), our proposed method utilizes the advanced contextual understanding and generative capabilities of the LLMs to effectively purify the adversarial examples. More concretely, 
we employ prompt engineering to leverage Large Language Models (LLMs) in the retrieval of purified examples from provided adversarial instances, ensuring both semantic similarity and accurate classification. Our novel method exhibits impressive performance across diverse classifiers, resulting in an average accuracy improvement of over 65\% under adversarial attacks.
%
\section{Acknowledgements}
This work is supported by Army Research Office (ARO) W911NF2110030 and Army Research Laboratory (ARL) W911NF2020124. Opinions, interpretations, conclusions, and recommendations are those of the authors' and
should not be interpreted as representing the official views or policies of the Army Research Office or the Army Research Lab.
\bibliographystyle{splncs04}
\bibliography{references}

\end{document}